\documentclass{article}

     \PassOptionsToPackage{numbers, compress}{natbib}

 \usepackage[preprint]{neurips_2020}
\usepackage[utf8]{inputenc} 
\usepackage[T1]{fontenc}    
\usepackage{hyperref}       
\usepackage{booktabs}       
\usepackage{amsfonts}       
\usepackage{nicefrac}       
\usepackage{microtype}      
\usepackage{graphicx}       
\usepackage{amsmath}        
\usepackage{multicol}       
\usepackage{caption}
\usepackage{subcaption}

\usepackage{comment}
\usepackage{color}

\newcommand{\hlam}[0]{ \lambda}
\newcommand{\xL}[0]{x_\text{L}}
\newcommand{\xR}[0]{x_\text{R}}

\title{Unsupervised Neural Networks for Quantum Eigenvalue Problems}

\author{%
  Henry Jin\\ 
  Department of Electrical Engineering and Computer Sciences, University of California, Berkeley\\ Berkeley, California 94720, United States \\
  \texttt{helinjin@berkeley.edu} 
\And
Marios Mattheakis
, Pavlos Protopapas \\
John A. Paulson School of Engineering and Applied Sciences, Harvard University \\
Cambridge, Massachusetts 02138, United States \\
\texttt{ \{mariosmat, pavlos\}@seas.harvard.edu}
}
\graphicspath{ {images/} }

\begin{document}

\maketitle

\begin{abstract}
Eigenvalue problems  are critical to several fields of science and engineering.  We present a novel  unsupervised neural network for discovering eigenfunctions and eigenvalues for  differential eigenvalue problems with solutions that identically satisfy the boundary conditions. A scanning mechanism is embedded allowing the method to find an arbitrary number of solutions. The network optimization is data-free and depends solely on the predictions. The unsupervised method  is used to solve the quantum infinite well and  quantum oscillator eigenvalue problems.
\end{abstract}

\section{Introduction}

Differential equations are prevalent in every field of science and engineering, ranging from physics to economics. Thus, extensive research has been done on developing numerical methods for solving differential equations.
With the unprecedented availability of computational power, neural networks hold promise in redefining how computational problems are solved. Among other applications, unsupervised neural networks are capable of  efficiently   solving  differential equations \cite{ourLib2020, pnas2018,  lagaris1998, nips2018,  mattheakis2020hamiltonian}.
These are unsupervised, data-free methods where the  optimization depends solely on the network predictions. 
The neural network solvers pose several advantages over numerical integrators: the obtained solutions are analytical and differentiable \cite{lagaris1998}, 
{networks are more robust to} the `curse of dimensionality' \cite{pnas2018},  numerical errors are not accumulated \cite{mattheakis2020hamiltonian}, and a family of solutions corresponding to different initial or boundary conditions can be constructed \cite{cedric2020}.

%

Differential eigenvalue equations with boundary conditions 
appear in a wide range of problems, including quantum mechanics and applied mathematics. 
Efficient numerical iterative methods, such as finite difference method,  have been developed  for solving eigenvalue problems, but they share the drawbacks common to all numerical integrators.
{Lagaris et al.} \cite{lagaris_eigen} have shown that neural networks are able to solve eigenvalue problems and proposed a  {  partially iterative method that solves a differential equation with a fixed eigenvalue at each iteration.} 
%
{Our contribution includes} a novel unsupervised neural network architecture that  simultaneously learns eigenvalues and the associated eigenfunctions using a scanning mechanism. 
The proposed technique is an extension to neural network differential equation solvers and, consequently, acquires all the benefits that  network solvers have over  numerical integrators. Moreover, our method has an additional advantage over integrators in that it discovers solutions that identically satisfy the boundary conditions.
We assess the performance of the proposed architecture by solving two  standard eigenvalue problems of quantum mechanics, namely, the infinite well, and the quantum harmonic oscillator.


\section{Methodology}
\label{gen_inst}

\begin{figure}[h]
    \centering
    \includegraphics[width = 0.7\textwidth]{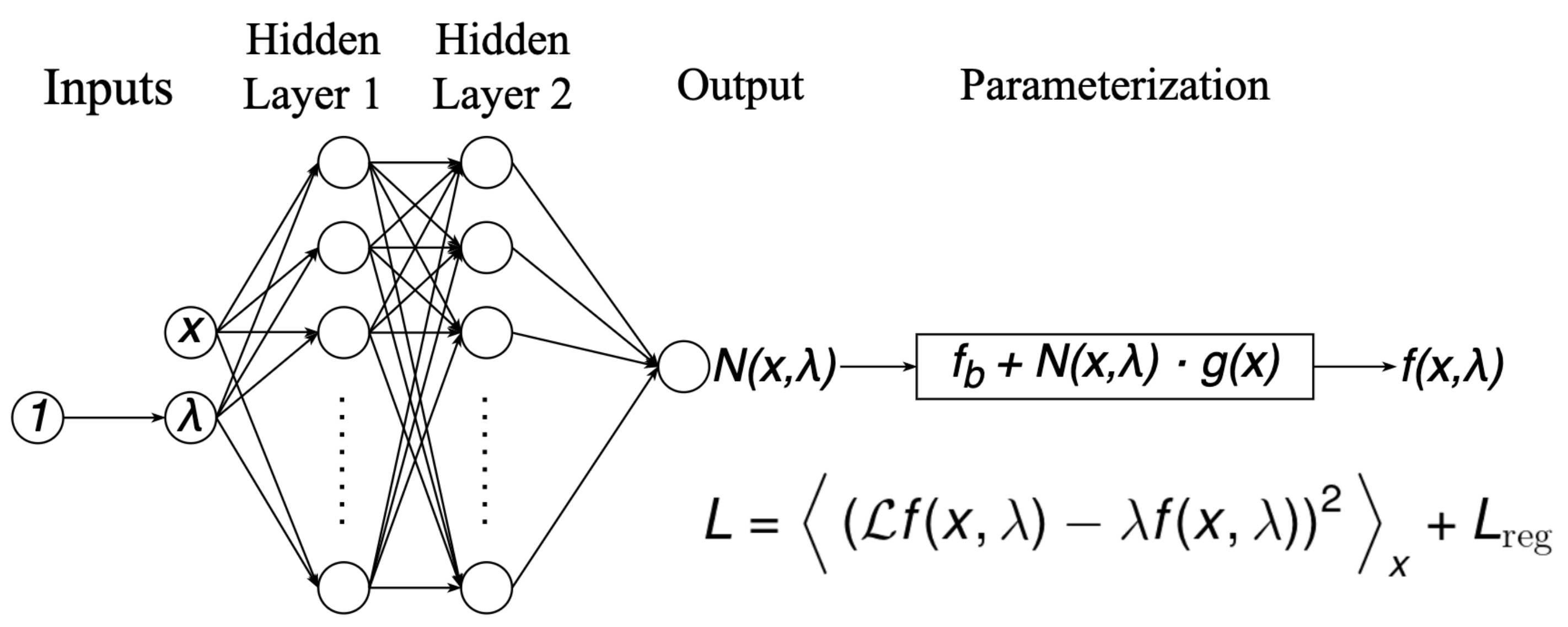}
    \caption{Adopted eigenvalue problem architecture. }
    \label{fig:finalarch}
\end{figure}

We consider an eigenvalue problem that exhibits the form:
\begin{align}
\label{eq:EigenProblem}
    \mathcal{L}f(x) = \lambda f(x),
\end{align}
where $x$ is the spatial variable, $\mathcal{L}$ is a differential operator which depends on $x$ and its derivatives, $f(x)$ is the eigenfunction, and $\lambda$ is the associated eigenvalue. We assume homogeneous  Dirichlet boundary conditions at $x_{L}$ and $x_{R}$ such that $f(\xL)=f(\xR)=f_b$, 
where $f_b$ is a constant boundary value.
For a given and fixed eigenvalue, Eq. (\ref{eq:EigenProblem}) is an  equation that  can be solved by neural network methods suggested in \cite{ourLib2020, lagaris_eigen,lagaris1998}.
We introduce a new  architecture shown in Fig. \ref{fig:finalarch}, which is capable of solving Eq. (\ref{eq:EigenProblem}) when both $f(x)$ and $\lambda$ are unknown.
The network  takes two inputs,  $x$ and  $1$. The constant input feeds a single linear neuron (affine transformation) that is 
updated through optimization, allowing  the network to find constant   $\hlam$. 
The $x$ and $\hlam$ feed a  feed-forward fully-connected network that returns an output function $N(x,\hlam)$. The predicted eigenfunctions $f(x,\hlam)$ is defined by using  a parametric trick, similar to Ref. \cite{mattheakis2020hamiltonian}, according to the equation:
\begin{align}
\label{eq:parSol}
    f(x,\hlam) = f_b + g(x)N(x,\hlam),
\end{align}
where we employ the parametric function 
\begin{align}
\label{eq:parametric}
    g(x) = \left(1-e^{-(x-\xL)}\right)\left(1-e^{-(x-\xR)}\right), 
\end{align}
which embeds the boundary conditions in the $f(x,\hlam)$.

Our  aim   is to discover pairs of $f(x, \hlam )$ and $\hlam$  that satisfy Eq. (\ref{eq:EigenProblem}). This is achieved by minimizing, during the optimization, a loss function $L$   defined by Eq. (\ref{eq:EigenProblem}) as:
\begin{align}
\label{eq:Loss}
    L &=  L_\text{DE} +  L_\text{reg} \nonumber \\
      &=  \Big \langle \big( \mathcal{L}f(x, \lambda) - \lambda f(x, \lambda) \big)^2  \Big \rangle_x +  L_\text{reg},
\end{align}
 where $\langle \cdot \rangle_x$ represents  averaging with respect to $x$. Any derivative with respect to  $x$ contained in  $\mathcal{L}$ is calculated by using the auto-differentiation technique \cite{Paszke2017AutomaticDI}.
The $L_\text{reg}$  in Eq. (\ref{eq:Loss}) contains regularization loss functions and is defined as: $L_\text{reg} = \nu_f L_f +\nu_\lambda L_\lambda + \nu_\text{drive} L_\text{drive}$. { Empirically, for the problems discussed below, we found  the optimal set  $\nu_ f= \nu_\lambda=\nu_\text{drive}=1$.} The $L_f$ and $L_\lambda$  are used to avoid learning  trivial eigenfunctions and eigenvalues respectively, while $ L_\text{drive}$  motivates the network to scan for higher eigenvalues, {as we explain below}. The regularization functions   are defined as:
\begin{align}
\label{eq:nontriv}
    L_{f} = \frac{1}{f(x, \lambda)^2}, \quad \quad \quad \quad 
    L_{\lambda} = \frac{1}{\lambda^2}, \quad \quad \quad \quad
    L_{\text{drive}} = e^{-\lambda + c}.
\end{align}
During the optimization, a scheduled scanning algorithm increases  the $L_\text{drive}$ by increasing $c$  in regular intervals. That forces the network to search for larger eigenvalues and the associated eigenfunctions.  At each interval the network is optimized and  the model parameters are stored when sufficiently low $L_\text{DE}$ is achieved. 
We emphasize that the loss function solely  depends  on the predictions of the network and, therefore, the training process is data-free, resulting in an unsupervised learning method.  
For the training, a batch of $x$ points in the interval $[\xL, \xR]$ is selected as input. In every training iteration (epoch) the input points are perturbed by a Gaussian noise \cite{mattheakis2020hamiltonian}. Adam optimizer is used \cite{adam} with a learning rate of $8 \cdot 10^{-3}$. 
%
We use two hidden layers of 50 neurons per layer with  trigonometric $\sin(\cdot)$  activation function.  The use of  $\sin(\cdot)$ instead of more common activation functions, such as Sigmoid$(\cdot)$ and $\tanh(\cdot)$, significantly accelerates the network's convergence to a solution \cite{mattheakis2020hamiltonian}.  We  implemented  the proposed neural network in pytorch  \cite{Paszke2017AutomaticDI} and published the code on github \footnote{https://github.com/henry1jin/eigeNN}.

\section{Experiments}
\label{headings}

We evaluate the effectiveness of the proposed method  by solving the eigenvalue problem defined by Schrodinger's equation. This is a fundamental equation in quantum mechanics that  describes the state wavefunction $\psi(x)$  and energy $E$ of a quantum system. We are interested in solving the one-dimensional stationary Schrodinger equation  defined as:
\begin{align}
\label{eq:QMeigen}
    \left[ -\frac{\hbar^2}{2m}\frac{\partial^2}{\partial x^2} + V(x)\right] \psi(x) &= E\psi(x),
\end{align}
where $\hbar$ and $m$ stand for Planck constant and mass {which}, without loss of  generality, can be set to   $\hbar=m=1$.   Equation (\ref{eq:QMeigen}) defines an eigenvalue problem where $\psi(x)$ and $E$ denote the {eigenfunction $f(x,\lambda)$ and eigenvalue $\lambda$ pair}. A  boundary value eigenvalue problem is defined by considering  a  potential function $V(x)$ and boundary conditions of $\psi(x)$.
We assess the performance of the proposed network architecture by solving Eq. (\ref{eq:QMeigen}) for the potential functions of the infinite square well  and the harmonic oscillator, both of which have known analytical solutions.

\subsection{Infinite Square Well}

\begin{figure}[h]
    \centering
    \includegraphics[scale=.25]{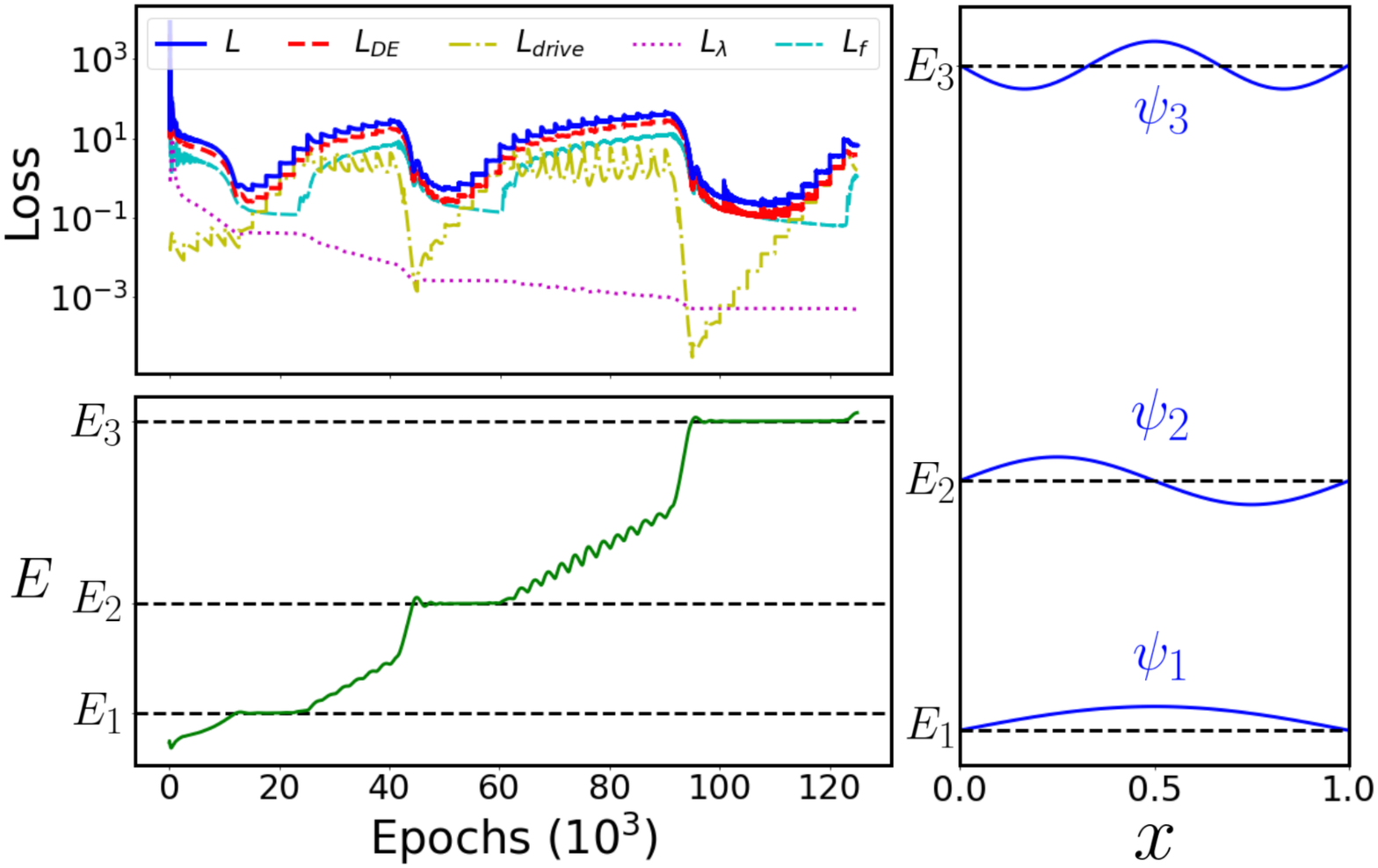}
    \caption{Infinite square well: Left panel shows the loss functions and the predicted energy during the training; dashed lines indicate the exact energy levels.  Right plot outlines the predicted eigenfunctions (blue) and eigenvalues (dashed black).  The errors are of the order $10^{-3}$  and $10^{-4}$ for $\psi$ and $E$.}
    \label{fig:infiniteplots}
\end{figure}

The infinite square well  problem is characterized by the following potential function:
\begin{align}
\label{eq:infpot}
    V(x) = \begin{cases} 
          0 & 0\leq x\leq \ell \\
          \infty & \text{otherwise}
       \end{cases},
\end{align}
 where  the length of the well is set to $\ell=1$. The exact  eigenfunctions and eigenvalues read
\begin{align}
\label{eq:infefunc}
\psi_n(x) = \begin{cases} 
          \sqrt{{2}} \sin({n\pi}x) & 0\leq x\leq 1 \\
          0 & \text{otherwise}
       \end{cases},
\quad \quad       E_n = \frac{n^2\pi^2}{2},
\end{align}
where $n$ is a positive integer and  indicates different solutions. 
The eigenfunctions are strictly zero outside of the well, implying the boundary conditions $\psi(0)=\psi(1)=0$.  The Eqs. (\ref{eq:parSol}) and (\ref{eq:parametric}) ensure the boundary conditions by setting $\xL=0$, $\xR=1$, and $f_b=0$. The proposed  scanning model is capable of solving for an arbitrary number of the first $n$ states. In Fig. \ref{fig:infiniteplots} we show results up to  $n=3$. 
The left panel presents the loss functions of Eqs. (\ref{eq:Loss}) and (\ref{eq:nontriv}) (upper), and the predicted $E$ (lower) during the network optimization.  
The scanning algorithm pushes the predicted eigenvalue upwards. The loss falls precipitously when an eigenfunction  is found, and the energy shows plateaus at these exact eigenvalues (indicated by dashed black line) of Eq. (\ref{eq:infefunc}). The loss function in Fig. \ref{fig:infiniteplots} depicts three dips, which correspond to three plateaus in the energy. This behavior gives a physical meaning to the loss function, since by inspecting  $L$ during the training we can draw the eigenstates. The right  panel  shows the extracted $\psi(x)$ (blue) and $E$ (dashed black) at each plateau. Comparing with the exact solutions of Eq. (\ref{eq:infefunc}), the order of magnitude of errors  are $10^{-3}$ and $10^{-4}$  for $\psi_n$ and $E_n$, respectively.

\subsection{Quantum Harmonic Oscillator}

The harmonic oscillator is characterized by the quadratic potential function  
\begin{align}
\label{eq:harmpot}  
V(x) = \frac{1}{2}kx^2,
\end{align}
where $k$ is the force constant and is considered to be $k=4$. The exact solutions for the eigenfunctions and energies are  given in terms of Hermite polynomials $H_n$ as 
\begin{align}
\label{eq:harmsols}
\psi_n(x) = \frac{1}{\sqrt{2^n n!}}  \frac{e^{-\frac{ x^2}{2}}}{\pi^{1/4}}     H_n\left( x \right), 
\quad \quad   E_n =   n + \frac{1}{2} .
\end{align}

The boundary conditions for the quantum oscillator problem  dictate the wavefunction to vanish at infinity, that is, $\psi(-\infty)=\psi(\infty)=0$. In numerical methods,  infinity is assumed to be a large number compared to the potential dimensions. We adopt the same approach and consider the boundary conditions $\psi(-6)=\psi(6)=0$. Thus, Eqs.  (\ref{eq:parSol}) and (\ref{eq:parametric}) ensure the boundary conditions by setting $\xL=-6$, $\xR=6$, and $f_b=0$.

\begin{figure}[h]
    \centering
    \includegraphics[scale=.26]{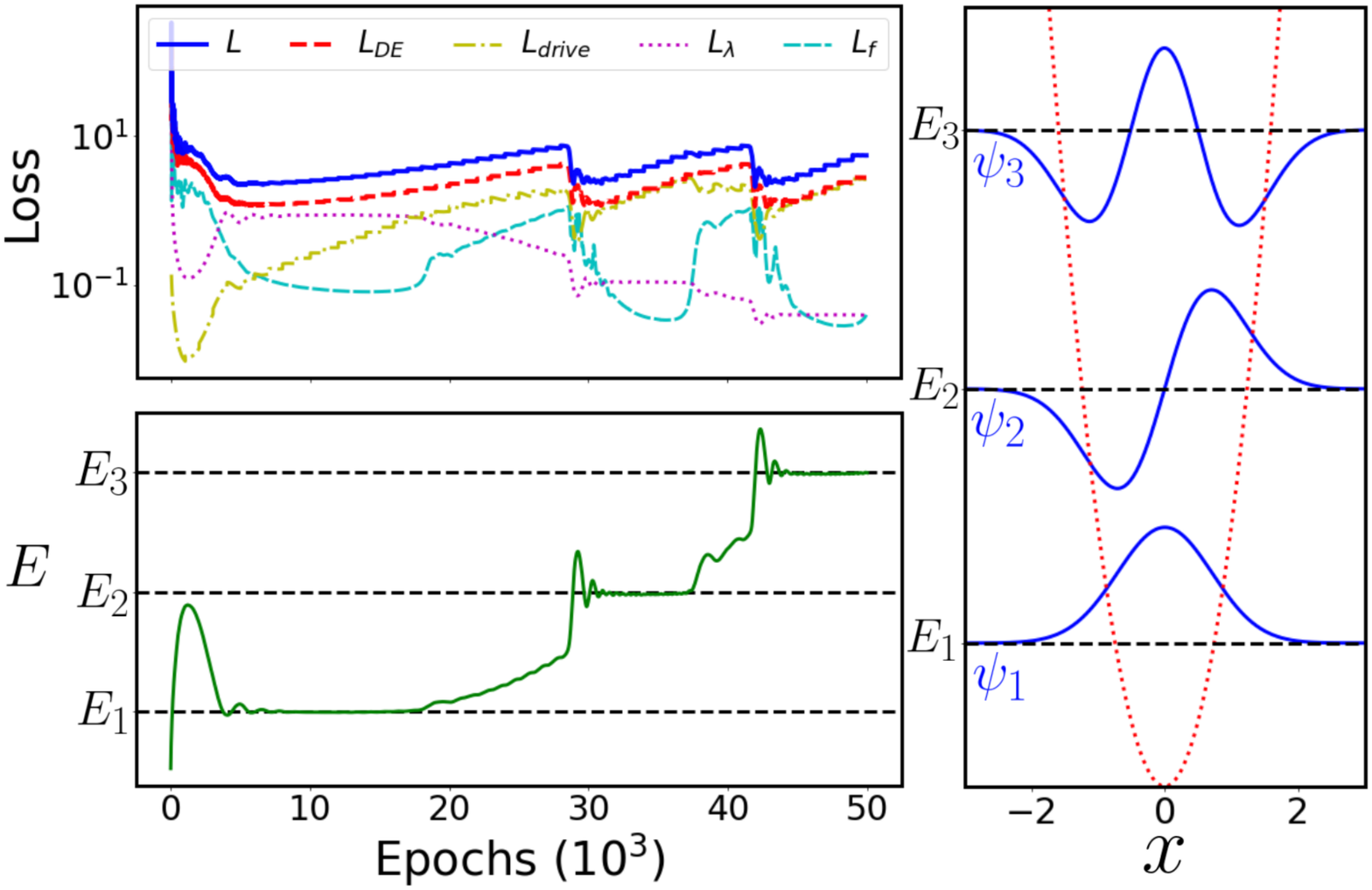}
    \caption{Quantum harmonic oscillator: Top left shows the various loss terms. Bottom left plots the history of the predicted energy. Right plot shows the eigenfunctions found by the model; the red dotted line outlines the potential. The order of magnitude of errors are $10^{-2}$ for $\psi$ and $10^{-2}$ for $E$.}
    \label{fig:harmonicplots}
\end{figure}

The proposed scanning neural network method is employed to discover the first three eigenstates for the quantum harmonic oscillator. The left panel in Fig. \ref{fig:harmonicplots} shows the drops in the total loss (blue line) that correspond to plateaus in the eigenvalue during training (green), indicating that an eigenvalue has been found. The predicted $\psi_n(x)$ and $E_n$ are presented in the right panel in Fig.  \ref{fig:harmonicplots} by solid blue and dashed black lines, respectively, while the red dotted curve outlines the potential energy. 


\section{Conclusion}
In recent years, there has been a growing interest in the application of  neural networks to study differential equations. In this work, we introduced a neural network  that is capable  of discovering  eigenvalues and eigenfunctions for boundary conditioned differential eigenvalue problems. The obtained solutions identically satisfy the given boundary conditions. A scanning mechanism allows the network to find an arbitrary number of eigenvalues and associated eigenfunctions. 
Inspecting the loss function during training allows one to draw the eigenstates,  providing  a physical meaning to the loss function.
The optimization solely depends on the network's predictions, consisting of an unsupervised learning method.  We  demonstrated the capability of the proposed architecture by solving  the   infinite well and harmonic oscillator quantum problems.

\begin{ack}
The authors would like to acknowledge fruitful discussions with Dr. David Sondak. 
\end{ack}


\bibliographystyle{plain}

\end{document}